# Theoretical aspects of a Novel Scaled Boundary Finite Element formulation in Computational Electromagnetics

V.S.Prasanna Rajan*, K.C.James Raju
School of Physics, University of Hyderabad, Hyderabad - 500 046, India

**Abstract :** The basic theory for a novel scaled boundary finite element formulation is developed for the general vector finite element functional in **H** formulation governing waveguides and resonators in electromagnetics. Also, the advantages of the proposed scaled boundary finite element formulation is explained.

**Introduction** : The Finite Element method has proved to be one of the most versatile technique to analyze and design microwave devices. In particular, the Vector Finite element method has been successfully applied for the analysis of complex three dimensional geometries occurring in electromagnetics. However, when the vector finite element method is applied to the three dimensional geometries, in general, it necessitates discretization in all the three dimensions. This in turn requires huge memory requirements for storing element datas like node numbers, local and global coordinates of elements, and connectivity data between elements in all the three dimensions. Also the discretization in three dimensions significantly increases the computation time for the eigen value and eigen vector computation of the resulting finite element equation. matrices can be exploited to simplify the eigen value and eigen vector computation.

___________________________________________________
∗ Corresponding author : vsprajan@yahoo.com , kcjrsprs@uohyd.ernet.in

In this paper, a novel scaled boundary finite element approach , which was initially developed by Chongmin Song and John.P.Wolf [1-13] to successfully solve elastodynamic and allied problems in civil engineering , is reformulated for the general functional governing cavities and waveguides in electromagnetics.

The scaled boundary finite element method is based entirely on finite elements but with a discretization only on the boundary. Unlike the boundary element method, this method doesn't require any fundamental solution (Green's function) to be known in advance. In order to apply this novel method, a scaling center is first chosen in such a way that the total boundary under consideration is visible from it. In case of geometries where it is not possible to find such a scaling center, the entire geometry is sub-structured [13] , and in each sub structure the scaling center can be chosen and the scaled boundary finite element method can be applied to each sub structure independently and can be combined together so that in effect, the whole geometry is analyzed. The concept of the scaled boundary finite element is described in detail in [1] but is repeated here in brief for convenience. By scaling the boundary in the radial direction with respect to the scaling center O with a scaling factor smaller than 1, the whole domain is covered.

This is shown in Fig.(1).

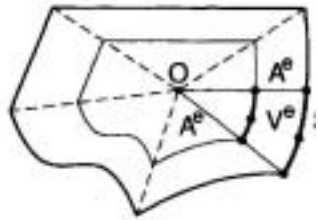

**Fig.1. Scaled Boundary (section)**

The scaling applies to each surface finite element. Its discretized surface on the boundary is denoted as $S^e$ (superscript e for element). Continuous scaling of the element yields a pyramid with volume $V^e$. The scaling center O is at its apex. The base of the pyramid is the surface finite element. The sides of the pyramid forming the boundary $A^e$ follow from connecting the curved edge of the surface finite element to the scaling center by straight lines. No discretization on $A^e$ occurs. Assembling all the pyramids by connecting their sides which corresponds to enforcing compatibility and equilibrium results in the total medium with volume V and the closed boundary S. No boundaries $A^e$ passing through the scaling center remain. Mathematically, the scaling corresponds to a transformation of the coordinates for each finite element, resulting in the two curvilinear local coordinates in the circumferential directions on the surface and the dimensionless radial coordinate representing the scaling factor. This transformation is unique due to the choice of the scaling center from which the total boundary of the geometry is visible. Summarizing, scaling of the boundary discretized with finite elements is applied, which explains the name of the scaled boundary finite-element method. The advantages of the scaled boundary finite element method are as follows :

a) Reduction of the spatial dimension by one, reducing the discretization effort and the number of degrees of freedom.
b) No fundamental solution required which permits general anisotropic material to be addressed and eliminates singular integrals.
c) Radiation condition at infinity satisfied exactly for unbounded media.
d) No discretization on that part of the boundary and interfaces between different materials passing through the scaling center.

e) Analytical expression for the field variables in the radial direction.

f) Converges to the exact solution in the finite-element sense in the circumferential directions.

g) Tangential Continuity conditions at the interfaces of different elements are automatically satisfied.

**Scaled boundary transformation of the geometry** [1]

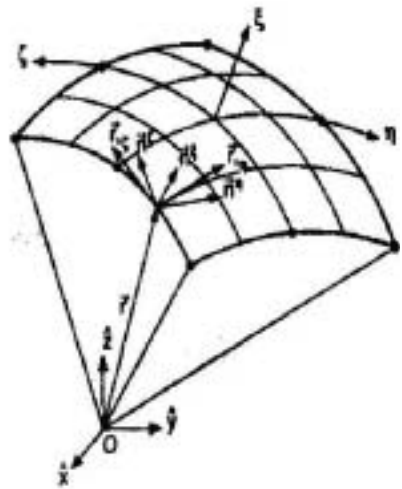

**Fig.2. Scaled boundary transformation of geometry of surface finite element**

The transformation of the geometry corresponding to the scaled boundary in Fig.1 is addressed. In Fig.2., given above, a specific finite element is addressed. The coordinates on the doubly-curved boundary are denoted as x,y,z. A point on the boundary is described by its position vector

**r** = x **i** + y **j** + z **k**  (1)

The cartesian coordinates $\hat{x}, \hat{y}, \hat{z}$ are transformed to the curvilinear coordinates $\xi, \eta, \zeta$. The scaling center is chosen in the interior of the domain. It coincides with the origin of the coordinate system $\hat{x}, \hat{y}, \hat{z}$. The dimensionless radial coordinate $\xi$ is measured from the scaling center along the position vector

$$\hat{\mathbf{r}} = \hat{x}\mathbf{i} + \hat{y}\mathbf{j} + \hat{z}\mathbf{k} \tag{2}$$

(unit vectors **i,j,k**), $\xi$ is constant (equal to 1) on the boundary. In a practical application, the geometry of the boundary is so general that only a piecewise description is feasible. (Doubly-curved) surface finite elements are used.

A specific finite element is addressed. The geometry of this finite element on the boundary is represented by interpolating its nodal coordinates $\{x\},\{y\},\{z\}$ using the local coordinates $\eta,\zeta$.

$$x(\eta,\zeta) = [N(\eta,\zeta)]\{x\} \tag{3a}$$

$$y(\eta,\zeta) = [N(\eta,\zeta)]\{y\} \tag{3b}$$

$$z(\eta,\zeta) = [N(\eta,\zeta)]\{z\} \tag{3c}$$

with the mapping functions

$$[N(\eta,\zeta)] = [N_1(\eta,\zeta)\ N_2(\eta,\zeta)\ \ldots\ldots] \tag{4}$$

A point in the domain is obtained by scaling that on the boundary.

$$\hat{\mathbf{r}} = \xi \mathbf{r} \tag{5}$$

Expressed in coordinates

$$\hat{x}(\xi,\eta,\zeta) = \xi x(\eta,\zeta) \tag{6a}$$

$$\hat{y}(\xi,\eta,\zeta) = \xi y(\eta,\zeta) \tag{6b}$$

$$\hat{z}(\xi,\eta,\zeta) = \xi z(\eta,\zeta) \tag{6c}$$

applies with $\xi=1$ on the boundary and $\xi=0$ in the scaling center. It is shown in detail in [1] that the relationship between the differential operators in the

$(\hat{x}, \hat{y}, \hat{z})$ and $(\xi, \eta, \zeta)$ can be expressed as

$$\begin{Bmatrix} \dfrac{\partial}{\partial \hat{x}} \\ \dfrac{\partial}{\partial \hat{y}} \\ \dfrac{\partial}{\partial \hat{z}} \end{Bmatrix} = \dfrac{g^\xi}{|J|}\{n^\xi\}\dfrac{\partial}{\partial \xi} + \dfrac{1}{\xi}\left(\dfrac{g^\eta}{|J|}\{n^\eta\}\dfrac{\partial}{\partial \eta} + \dfrac{g^\zeta}{|J|}\{n^\zeta\}\dfrac{\partial}{\partial \zeta}\right) \qquad (7)$$

where
$$|J| = x(y_{,\eta}z_{,\zeta} - z_{,\eta}y_{,\zeta}) + y(z_{,\eta}x_{,\zeta} - x_{,\eta}z_{,\zeta}) + z(x_{,\eta}y_{,\zeta} - y_{,\eta}x_{,\zeta}) \qquad (8)$$

$$g^\xi = |\mathbf{g}^\xi| = |\mathbf{i}(y_{,\eta}z_{,\zeta} - z_{,\eta}y_{,\zeta}) + \mathbf{j}(z_{,\eta}x_{,\zeta} - x_{,\eta}z_{,\zeta}) + \mathbf{k}(x_{,\eta}y_{,\zeta} - y_{,\eta}x_{,\zeta})| \qquad (9)$$

$$g^\eta = |\mathbf{g}^\eta| = |\mathbf{i}(zy_{,\zeta} - yz_{,\zeta}) + \mathbf{j}(xz_{,\zeta} - zx_{,\zeta}) + \mathbf{k}(yx_{,\zeta} - xy_{,\zeta})| \qquad (10)$$

$$g^\zeta = |\mathbf{g}^\zeta| = |\mathbf{i}(yz_{,\eta} - zy_{,\eta}) + \mathbf{j}(zx_{,\eta} - xz_{,\eta}) + \mathbf{k}(xy_{,\eta} - yx_{,\eta})| \qquad (11)$$

In all the above expressions the subscripts $\xi$ $\eta$ $\zeta$ indicate the partial derivatives with respect to the subscripted variables and $\{n^\xi\}$, $\{n^\eta\}$, $\{n^\zeta\}$ are the matrices containing the $\xi$ $\eta$ and $\zeta$ components of the unit outward normal vectors to the surfaces $(\eta,\zeta)$, $(\zeta,\xi)$, and $(\xi,\eta)$ respectively on the boundary where $\xi = 1$ and

$$n^\xi = |\mathbf{n}^\xi| = |\mathbf{g}^\xi| / g^\xi \qquad (12)$$

$$n^\eta = |\mathbf{n}^\eta| = |\mathbf{g}^\eta| / g^\eta \qquad (13)$$

$$n^\zeta = |\mathbf{n}^\zeta| = |\mathbf{g}^\zeta| / g^\zeta \qquad (14)$$

The differential volume element dV in the $(\xi,\eta,\zeta)$ coordinate system is given by

$$dV = \xi^2 |J| d\xi\, d\eta\, d\zeta \qquad (15)$$

It is important to note that **| J | is evaluated only on the boundary**.

**Representation of the electromagnetic functional in the scaled boundary coordinates and the derivation of the scaled boundary finite element equation :**

The governing electromagnetic functional in **H** formulation is given by

$$\Pi = \int_\Omega \left(\nabla \times \mathbf{H} \cdot \varepsilon_r^{-1} \nabla \times \mathbf{H} - k_o^2 \mathbf{H} \cdot \mu_r \mathbf{H}\right) d\Omega \tag{16}$$

In this paper, the case where $\mu_r=1$ and $\varepsilon_r$ is constant and isotropic is addressed.

where $\mu_r$ is the relative permeability and $\varepsilon_r$ is the relative permittivity of the medium respectively. $\Omega$ denotes the domain under consideration. In the $(\xi,\eta,\varsigma)$ coordinates, the components of $\mathbf{H}$ are represented as

$$H_\xi = f_1(\xi) \sum_{i=0}^{m} \sum_{j=0}^{n} h_{\xi(i,j)} h_{\xi i}(\eta) h_{\xi j}(\zeta) \tag{17a}$$

$$H_\eta = f_2(\xi) \sum_{i=0}^{m} \sum_{j=0}^{n} h_{\eta(i,j)} h_{\eta i}(\eta) h_{\eta j}(\zeta) \tag{17b}$$

$$H_\zeta = f_3(\xi) \sum_{i=0}^{m} \sum_{j=0}^{n} h_{\zeta(i,j)} h_{\zeta i}(\eta) h_{\zeta j}(\zeta) \tag{17c}$$

where the functions $f_1$, $f_2$ and $f_3$ are unknown radial functions depending on the radial coordinate $\xi$, $h_{\xi(i,j)} h_{\eta(i,j)} h_{\zeta(i,j)}$ are unknown coefficients, and $h_i(\eta), h_j(\zeta)$ are the single variable functions of $\eta,\zeta$ representing the variations in $\varsigma$ and æ respectively and $m < n$ and $m \neq 0$ and $n \neq 0$

The form of the functions $h_i(\eta)$ and $h_j(\varsigma)$ are given as [14]
$h_o(r)=1-r$ (18a)

$h_1(r)=1+r$ (18b)

$h_i(r)=(1-r^2)r^{i-2}$ for i greater than or equal to 2. (18c)

Hence, $\vec{H}(\xi,\eta,\zeta) = H_\xi \mathbf{n}^{\hat{\xi}} + H_\eta \mathbf{n}^{\varsigma} + H_\zeta \mathbf{n}^{\varsigma}$ (18d)

It is shown in [14] that for two dimensional surface finite elements employing the functions of the form given above, the condition m<n is necessary in the double summation series expansion and only the tangential continuity between adjacent elements need to be imposed in order to avoid spurious modes, the spurious modes being the modes giving non zero divergence to **H**. The functions $f_1(\xi), f_2(\eta), f_3(\varsigma)$ are represented in the form of power series expansion in $\xi$.

$$f_1(\xi) = \sum_{k=0}^{N} a_k \xi^k \qquad (19a)$$

$$f_2(\xi) = \sum_{k=0}^{N} b_k \xi^k \qquad (19b)$$

$$f_3(\xi) = \sum_{k=0}^{N} c_k \xi^k \qquad (19c)$$

where **N=((m+1)(n+1)) −1** and $a_k$, $b_k$, $c_k$ are arbitrary unknown coefficients.

Since $a_k$, $b_k$, $c_k$ are arbitrary, they can be replaced by $h_{\xi(i,j)}$, $h_{\eta(i,j)}$, $h_{\varsigma(i,j)}$ respectively in the same sequence as the $h_{(i,j)}$ coefficients appear in the double summation series. The number of unknown coefficients in the double summation series for every single component of **H** is chosen to be equal to the number of unknown coefficients in the corresponding radial expansion. This results in the expression of N in terms of m and n given above. The effect of this replacement makes the radial expansion also in terms of the unknown $h_{(i,j)}$ coefficients. This has the advantage that the resulting finite element equation contains only a single type of unknown $h_{(i,j)}$ coefficients which can be solved for numerically. It is shown in [15] that the imposition of the condition div**H**=0 ($\mu_r$=1) is equivalent to the following set of constraint equations on the unknown $h_{(i,j)}$ coefficients.

$$\sum_{i=0}^{m}\sum_{j=0}^{n} (h_{\xi(i,j)} h_{\xi(0,0)})\left[(k_2 + k_3)_{i,j}\right] + (h_{\eta(i,j)} h_{\eta(0,0)})\left[(k_5 + k_6)_{i,j}\right] + (h_{\varsigma(i,j)} h_{\varsigma(0,0)})\left[(k_8 + k_9)_{i,j}\right] = 0$$

$$\text{for } k = 0 \quad ....(20a)$$

$$\sum_{i=0}^{m}\sum_{j=0}^{n} (h_{\xi(i,j)} h_{\xi(i_k,j_k)})\left[(k \cdot k_{1(i,j)}) + (k_2 + k_3)_{i,j}\right] + (h_{\eta(i,j)} h_{\eta(i_k,j_k)})\left[(k \cdot k_{4(i,j)}) + (k_5 + k_6)_{i,j}\right]$$
$$+ (h_{\varsigma(i,j)} h_{\varsigma(i_k,j_k)})\left[(k \cdot k_{7(i,j)}) + (k_8 + k_9)_{i,j}\right] = 0 \qquad \text{for } k > 0 \quad ....(20b)$$

where $h_{\xi(0,0)}$, $h_{\eta(0,0)}$ and $h_{\varsigma(0,0)}$ correspond to the unknown h coefficients for i=j=0 for

$H_\xi$, $H_\eta$, $H_\varsigma$ respectively.

$h_{\hat{1}(i,j_k)}, h_{\varsigma(i,j_k)}, h_{æ(i,j_k)}$ correspond to the unknown h - coefficients with corresponding (i, j) values for given $k > 0$.

The constants ($k_1$ to $k_9$) $_{(i,j)}$ are evaluated for every surface finite element and their expressions are given as follows.

$$k_{1(i,j)} = \int_{(\eta_1,\zeta_1)}^{(\eta_2,\zeta_2)} \frac{g^\xi}{|J|} h_{\xi i}(\eta) h_{\xi j}(\zeta) d\eta d\zeta \tag{21a}$$

$$(k_2 + k_3)_{i,j} = \int_{(\eta_1,\zeta_1)}^{(\eta_2,\zeta_2)} \frac{1}{|J|} [(g^\eta n_x^\eta h'_{\xi i}(\eta) h_{\xi j}(\zeta)) + (g^\zeta n_x^\zeta h_{\xi i}(\eta) h'_{\xi j}(\zeta))] d\eta d\zeta \tag{21b}$$

$$(k_4)_{i,j} = \int_{(\eta_1,\zeta_1)}^{(\eta_2,\zeta_2)} \frac{1}{|J|} (g^\xi n_y^\xi h_{\eta i}(\eta) h_{\eta j}(\zeta)) d\eta d\zeta \tag{21c}$$

$$(k_5 + k_6)_{i,j} = \int_{(\eta_1,\zeta_1)}^{(\eta_2,\zeta_2)} \frac{1}{|J|} [(g^\eta n_y^\eta h'_{\eta i}(\eta) h_{\eta j}(\zeta)) + (g^\zeta n_y^\zeta h_{\eta i}(\eta) h'_{\eta j}(\zeta))] d\eta d\zeta \tag{21d}$$

$$(k_7)_{i,j} = \int_{(\eta_1,\zeta_1)}^{(\eta_2,\zeta_2)} \frac{1}{|J|} (g^\xi n_z^\xi h_{\varsigma i}(\eta) h_{\varsigma j}(\zeta)) d\eta d\zeta \tag{21e}$$

$$(k_8 + k_9)_{i,j} = \int_{(\eta_1,\zeta_1)}^{(\eta_2,\zeta_2)} \frac{1}{|J|} [(g^\eta n_z^\eta h'_{\varsigma i}(\eta) h_{\varsigma j}(\zeta)) + (g^\zeta n_z^\zeta h_{\varsigma i}(\eta) h'_{\varsigma j}(\zeta))] d\eta d\zeta \tag{21f}$$

In all the above expressions(21a-21f), h' denotes the derivative of h with respect to the variable in the curved bracket. The subscripts denote the respective component terms of **H**($\xi,\eta,\varsigma$). The upper and lower limits of $\eta$ and $\varsigma$ corresponds to their limits for every individual surface finite element. The constraint equations (20 a,b) generate a total of N+1 constraints for every surface finite element. These constraints are to be necessarily imposed so that the spurious modes are eliminated. The scaled boundary finite element equation is formulated as follows. The expressions for **H** in terms of its components in ($\xi,\eta,\varsigma$) can be concisely written as

$H_\xi = f_1(\xi)g_1(\eta,\varsigma)$                             (22a)

$H_\eta = f_2(\xi)g_2(\eta,\varsigma)$                             (22b)

$H_\varsigma = f_3(\xi)g_3(\eta,\varsigma)$                             (22c)

Rewriting (16) using (7) and (22) for $\mu_r=1$ we get for a single surface element,

$$\int_0^1 \int_{(\eta,\varsigma)} \left[ \frac{\varepsilon_r^{-1}}{|J|} \right] \left[ ((\xi g^\xi)^2 (A+B+C)) + ((g^\eta)^2 (D+E+F)) + ((g^\varsigma)^2 (G+H+I)) \right]$$
$$+ \frac{2\varepsilon_r^{-1}}{|J|} \left[ ((g^\eta g^\varsigma)(J+K+L)) + ((g^\xi g^\eta)(M+N+P)) + ((g^\xi g^\varsigma)(Q+R+S)) \right]$$
$$- k_0^2 \left[ (f_1(\xi)g_1(\eta,\varsigma))^2 + (f_2(\xi)g_2(\eta,\varsigma))^2 + (f_3(\xi)g_3(\eta,\varsigma))^2 \right] d\xi d\eta d\varsigma = 0$$

....(23)

The expressions for the terms A to S given in the integral have the following mathematical structure. These expressions involve the products of the terms involving the derivatives of **H** and the normal vectors which are given as,

$$A = \left[ n_y^\xi \frac{\partial H_\zeta}{\partial \xi} - n_z^\xi \frac{\partial H_\eta}{\partial \xi} \right]^2 \qquad (24a)$$

$$B = \left[ n_z^\xi \frac{\partial H_\varepsilon}{\partial \xi} - n_x^\xi \frac{\partial H_\zeta}{\partial \xi} \right]^2 \qquad (24b)$$

$$C = \left[ n_x^\xi \frac{\partial H_\eta}{\partial \xi} - n_y^\xi \frac{\partial H_\xi}{\partial \xi} \right]^2 \qquad (24c)$$

$$D = \left[ n_y^\eta \frac{\partial H_\zeta}{\partial \eta} - n_z^\eta \frac{\partial H_\eta}{\partial \eta} \right]^2 \qquad (24d)$$

$$E = \left[ n_z^\eta \frac{\partial H_\xi}{\partial \eta} - n_x^\eta \frac{\partial H_\zeta}{\partial \eta} \right]^2 \qquad (24e)$$

$$F = \left[ n_x^\eta \frac{\partial H_\eta}{\partial \eta} - n_y^\eta \frac{\partial H_\xi}{\partial \eta} \right]^2 \qquad (24f)$$

$$G = \left[ n_y^\zeta \frac{\partial H_\zeta}{\partial \zeta} - n_z^\zeta \frac{\partial H_\eta}{\partial \zeta} \right]^2 \qquad (24g)$$

$$H = \left[ n_z^\zeta \frac{\partial H_\xi}{\partial \zeta} - n_x^\zeta \frac{\partial H_\zeta}{\partial \zeta} \right]^2 \qquad (24h)$$

$$I = \left[ n_x^\zeta \frac{\partial H_\eta}{\partial \zeta} - n_y^\zeta \frac{\partial H_\xi}{\partial \zeta} \right]^2 \qquad (24i)$$

$$J = \left[ n_y^\eta \frac{\partial H_\zeta}{\partial \eta} - n_z^\eta \frac{\partial H_\eta}{\partial \eta} \right]\left[ n_y^\zeta \frac{\partial H_\zeta}{\partial \zeta} - n_z^\zeta \frac{\partial H_\eta}{\partial \zeta} \right] \qquad (24j)$$

$$K = \left[ n_z^\eta \frac{\partial H_\xi}{\partial \eta} - n_x^\eta \frac{\partial H_\zeta}{\partial \eta} \right]\left[ n_z^\zeta \frac{\partial H_\xi}{\partial \zeta} - n_x^\zeta \frac{\partial H_\xi}{\partial \zeta} \right] \qquad (24k)$$

$$L = \left[ n_x^\eta \frac{\partial H_\eta}{\partial \eta} - n_y^\eta \frac{\partial H_\xi}{\partial \eta} \right]\left[ n_x^\zeta \frac{\partial H_\eta}{\partial \zeta} - n_y^\zeta \frac{\partial H_\xi}{\partial \zeta} \right] \qquad (24L)$$

$$M = \left[ n_y^\xi \frac{\partial H_\zeta}{\partial \xi} - n_z^\xi \frac{\partial H_\eta}{\partial \xi} \right]\left[ n_y^\eta \frac{\partial H_\zeta}{\partial \eta} - n_z^\eta \frac{\partial H_\eta}{\partial \eta} \right] \qquad (24\,m)$$

$$N = \left[ n_z^\xi \frac{\partial H_\xi}{\partial \xi} - n_x^\xi \frac{\partial H_\zeta}{\partial \xi} \right]\left[ n_z^\eta \frac{\partial H_\xi}{\partial \eta} - n_x^\eta \frac{\partial H_\zeta}{\partial \eta} \right] \qquad (24\,n)$$

$$P = \left[ n_x^\xi \frac{\partial H_\eta}{\partial \xi} - n_y^\xi \frac{\partial H_\xi}{\partial \xi} \right] \left[ n_x^\eta \frac{\partial H_\eta}{\partial \eta} - n_y^\eta \frac{\partial H_\xi}{\partial \eta} \right] \quad (24\,p)$$

$$Q = \left[ n_y^\zeta \frac{\partial H_\zeta}{\partial \zeta} - n_z^\zeta \frac{\partial H_\eta}{\partial \zeta} \right] \left[ n_y^\xi \frac{\partial H_\zeta}{\partial \xi} - n_z^\xi \frac{\partial H_\eta}{\partial \eta} \right] \quad (24\,q)$$

$$R = \left[ n_z^\zeta \frac{\partial H_\xi}{\partial \zeta} - n_x^\zeta \frac{\partial H_\zeta}{\partial \zeta} \right] \left[ n_z^\xi \frac{\partial H_\xi}{\partial \xi} - n_x^\xi \frac{\partial H_\zeta}{\partial \xi} \right] \quad (24\,r)$$

$$S = \left[ n_x^\xi \frac{\partial H_\eta}{\partial \xi} - n_y^\xi \frac{\partial H_\xi}{\partial \xi} \right] \left[ n_x^\zeta \frac{\partial H_\eta}{\partial \zeta} - n_y^\zeta \frac{\partial H_\xi}{\partial \zeta} \right] \quad (24\,s)$$

The derivative terms in the above integrals can be evaluated from (17) and (22). The other terms not involving the derivatives can be evaluated using (17). In the equation (23), the integration with respect to $\xi$ from 0 to 1 can be performed independent of $\eta$ and $\zeta$ variables by treating them as constants with respect to $\xi$. The integrand of the collected terms of $\xi$ is a polynomial in $\xi$ with the h coefficients being the unknown constants. The net effect of the integration with respect to $\xi$ gives an equation similar to (23) but involving only the surface integrals containing the products of h coefficients coupled with some numerical constants. Then the constraint equations (20 a , 20 b) are imposed. This is done by elimination of common h coefficients from (20) and substituting the resultant in the equation containing only the surface integrals got after integrating with respect to $\xi$.

The net resultant equation obtained after the above procedure written in terms of a matrix equation after assembling together the individual finite element matrices by **imposing only tangential continuity between adjacent elements on the boundary** for every surface finite element is of the form[14]

$$\mathbf{A}h + k_0^2 \mathbf{B}h = 0 \qquad (25)$$

where h is the vector containing the unknown coefficients. The above eigen value equation can be can be solved by using standard numerical procedures. **It is important to note that equation (25) contains terms involving only the surface integrals even for the general 3-D case instead of the volume integrals, unlike the conventional finite element eigen value equation involving 3-D structures. This is the crucial advantage of the novel scaled boundary finite element method.**

**Conclusion :** A novel scaled boundary finite element formulation is introduced for the general electromagnetic functional governing wave guides and resonators. This method contains reduced set of unknowns as compared to conventional finite element equations. Also the discretization involved in this new formulation is only on the surface of the geometry for three dimensional case which is an enormous benefit in terms of resources and time for the eigen value and eigen vector computation for complex geometries.

**Acknowledgement :** The first author thanks Dr.John.P.Wolf of the Department of Civil Engg, Institute of Hydraulics and Energy, Swiss Federal Institute of Technology Lausanne,Switzerland for his crucial help in sending his research papers on the scaled boundary finite element method and for his helpful suggestions. The author also thanks the Council for scientific and Industrial Research (CSIR), New Delhi , India for providing the financial assistance in the form of Senior Research Fellowship in the research project sponsored by it.